\documentclass[letterpaper,twocolumn,10pt]{article}
\newcommand{\systemname}{\sc{Solar}}
\newcommand{\lll}{\mathrel{{=}{\llbracket}}}
\newcommand{\rrr}{\mathrel{{\rrbracket}{\Rightarrow}}}

\newcommand{\sourcecode}[1]{\texttt{#1}}

\usepackage{usenix2019}
\usepackage{tikz}
\usepackage{amsmath} 
\usepackage{pdflscape}
\usepackage{graphicx}
\usepackage[normalem]{ulem}
\usepackage[frozencache=true]{minted}
\usepackage[figure,vlined,linesnumbered]{algorithm2e}
\usepackage{stmaryrd}
\usepackage{mathtools}
\usepackage{amsfonts}
\usepackage{microtype}
\usepackage{etoolbox}
\apptocmd{\thebibliography}{\raggedright}{}{}

\graphicspath{{./figures}}
\setminted{
frame=lines,
baselinestretch=0.7,
fontsize=\footnotesize,
linenos,
baselinestretch=1,
breaklines,
xleftmargin=17pt
}
\SetKw{kwGoTo}{go to}
\SetKw{kwThrow}{throw}
\begin{document}

\date{}

\title{\Large \bf Detecting Standard Violation Errors in Smart Contracts} 

\author{
{\rm Ao Li}\\
University of Toronto\\
leo@cs.toronto.edu
\and
{\rm Fan Long}\\
University of Toronto\\
fanl@cs.toronto.edu
} 

\maketitle

\thispagestyle{empty}
\subsection*{Abstract}

We present {\systemname}, a new analysis tool for automatically detecting
standard violation errors in Ethereum smart contracts. Given the Ethereum
Virtual Machine (EVM) bytecode of a smart contract and a user specified
constraint or invariant derived from a technical standard such as ERC-20,
{\systemname} symbolically executes the contract, explores all possible
execution paths, and checks whether it is possible to initiate a sequence of
malicious transactions to violate the specified constraint or invariant.  Our
experimental results highlight the effectiveness of {\systemname} in finding
new errors in smart contracts. Out of the evaluated 779 ERC-20 and 310 ERC-721
smart contracts, {\systemname} found 255 standard violation errors in 197
vulnerable contracts with only three false positives. 237 out of the 255 errors
are zero-day errors that are not reported before. Our results sound the alarm
on the prevalence of standard violation errors in critical smart contracts that
manipulate publicly traded digital assets.

\section{Introduction}

Following the success of the cryptocurrencies~\cite{Nakamoto_bitcoin:a,
    wood2014ethereum}, blockchain has recently evolved into a technology
platform that powers secure, decentralized, and consistent transaction ledgers
at Internet-scale. 
Smart contract deployment is one of the most important features of many new
blockchain systems such as Ethereum~\cite{wood2014ethereum}.  Users can develop
smart contracts at a high level programming language to encode arbitrarily
complicated transaction rules.  These developed contracts are compiled to a low
level Ethereum Virtual Machine (EVM) bytecode and are eventually deployed to
Ethereum.  The contracts and the transaction rules are then faithfully executed
and enforced by all participants of the Ethereum network, eliminating any
potential counter-party risk of the encoded transactions in future.

Because smart contracts are in fact programs that directly manipulate critical
information such as digital assets, financial records, and even identities,
ensuring the correctness of these contracts is essential.
Unfortunately, human programmers often make mistakes and errors and the
consequence of programming errors in smart contracts is particularly severe.
For example, an anonymous attacker exploited an integer overflow vulnerability
of the BeautyChain smart contract in Ethereum to illegally generate a massive
amount of BECTokens, the underlying digital assets of the contract. This attack
caused the market cap of BECTokens, which was two billion dollars at the time
of the attack, to evaporate in days~\cite{bec}. Another famous example is the
DAO attack which caused 50 million dollars in losses and a community dividing
hard fork of Ethereum to recover the stolen fund~\cite{dao}.

On one hand, to alleviate the correctness problem of smart contracts, the
blockchain community has created many technical standards for common kinds of
smart contracts such as digital assets, identity tokens, and domain name
services~\cite{ERC137, ERC1484,ERC20, ERC721}.  A technical standard typically
defines a set of interface functions that a contract following the standard
should implement, together with specifications for each of the interface
functions.  This community driven effort is partially successful. For example,
most smart contracts that manage digital assets are now developed under
relevant technical standards such as ERC-20~\cite{ERC20} and
ERC-721~\cite{ERC721}.  However, it is unclear how many of such deployed
contracts actually conform to the standards, because a programming error may
still cause a contract implementation to deviate from its intended behavior.

On the other hand, previous research has been focusing on developing analysis
tools for Ethereum smart contracts to detect potentially dangerous low-level runtime errors such as integer
overflows and contract reentrance~\cite{mythril, trailofbits-manticore}.
However, these tools often report an excessive
amount of false warnings because 1) many reported low-level warnings are
actually benign errors and 2) these analysis tools do not accurately handle
storage access instructions in Ethereum Virtual Machine (EVM).  Also these
tools have a limited scope and cannot detect logic errors where the
implementation simply deviates from the intended specifications without any
runtime error.  Furthermore most of these tools require the access of the smart
contract source code and cannot apply to the EVM bytecode directly. These
drawbacks limit the applications of any prior art.

\subsection{{\systemname}}

We present {\systemname}, a novel symbolic execution analysis tool for
detecting standard violation errors in Ethereum smart contracts.  Given a set
of constraints and invariants derived from the standard to which a contract
conforms, {\systemname} validates whether it is possible for the contract to
violate the given constraints and invariants. If so, {\systemname} generates an
error report that include the initial contract state and the sequence of
transactions to trigger the violation.

Instead of focusing on low-level runtime errors, {\systemname} takes advantages
of the standardization effort by the blockchain community.  {\systemname}
exploits the fact that smart contracts following a common technical standard
such as ERC-20 are essentially sharing the same set of specifications defined
by the standard.  {\systemname} therefore provides an expressive language to
allow users to develop invariants and safety constraints derived from a
technical standard.  Once developed, these invariants and constraints can then
be applied to check all smart contracts that conform to the same standard.
Unlike previous tools, {\systemname} does not suffer from the problem of benign
errors, because all errors that {\systemname} detects correspond to deviations
between a contract implementation and the corresponding standard.

{\systemname} models the behavior of a smart contract as a state machine where
external transactions invoke associated contract functions to drive the state
transition. Given a constraint of a sequence of transactions, {\systemname}
symbolically executes the associated contract functions.  During the symbolic
execution, {\systemname} queries an SMT solver to determine whether there is a
possible assignment scheme of the initial contract state and the transaction
input parameters to cause the contract execution violating the specified
constraint.  

To build a symbolic execution engine for EVM bytecode, it is important to
accurately and efficiently handle storage load and store instructions.  There
are two kinds of storages in EVM, a persistent contract state that contains
non-volatile data across different transactions and a volatile memory that
contains temporary data during the process of one transaction.
One challenge {\systemname} faces is that load and store instructions of the
contract persistent state in EVM are often paired with special cryptographic hash
instructions to compute the address locations of the loaded/stored object.
Naively processing these cryptographic computations would generate complicated
symbolic expressions that SMT solvers cannot solve. 
Another challenge {\systemname} faces is that the bit width of the stored
volatile memory values is different from other parts of EVM.  The volatile
memory operates with 8-bit values, while the stack and the persistent state all
operate with 256-bit values.  Naively processing load and store instructions of
the volatile memory would generate a large number of bit extraction and
concatenation operations in symbolic expressions and would significantly slow
down the performance of the symbolic execution engine.

To address the challenge of cryptographic computations associated with persistent
state loads and stores, {\systemname} operates with a customized addressing
mechanism for the persistent state when running the symbolic
execution. This customized mechanism enables {\systemname} to replace cryptographic
computations with lightweight computations. In the same time, it still
guarantees the free of address collisions so that the program remains
functionally equivalent.  To address the challenge introduced by the volatile
memory bit width, {\systemname} uses a symbolic cache to track all 256-bit
values stored into the volatile memory. For the common case where the 256-bit
value is later loaded from the volatile memory as a whole, the cache will
detect this and return the cached value to skip processing bit extraction and
concatenation instructions.  These two techniques together enable efficient and
accurate symbolic execution for EVM bytecode.


\subsection{Experimental Results}

We evaluated {\systemname} with a benchmark set of the top 779 ERC-20 and the
top 310 ERC-721 smart contracts from Etherscan~\cite{etherscan}, the most
popular Ethereum blockchain explorer. ERC-20 and ERC-721 are two important
smart contract standards which define the specifications for implementing
fungible and non-fungible digital assets. Many of the evaluated contracts
manage digital assets that are publicly traded on crypto-exchanges.
We applied {\systemname} to check whether these contracts satisfy the total
supply invariant and the transfer functionalities defined by the ERC-20 and
ERC-721 standards.  

Our experimental results highlight the effectiveness of {\systemname} in
finding security errors in Ethereum smart contracts.  {\systemname} found in
total 255 standard violation errors in these 1089 contracts.  Note that 237 out
of the found 255 errors are zero-day errors, i.e., to the best of our
knowledge, we believe that these errors are not reported before.  We compared
the results of {\systemname} with Mythril~\cite{mythril}, the state-of-the-art
contract analysis tool developed by the Ethereum community.  For the same set
of smart contracts, Mythril reported 1115 high severity errors from 512 ERC-20
contracts and 595 errors from 217 ERC-721 contracts. However, our manual
analysis show that most of these reported errors are false positives and/or
benign errors. We sampled 100 smart contracts to manually compare the results
of {\systemname} and Mythril.  Among the 100 sampled contracts, {\systemname}
reported 36 errors from 24 contracts with no false positive, while Mythril
reported 127 errors from 60 contracts with only one true positive (which is
also found by {\systemname} as well)\footnote{64 of the Mythril reported errors
    in the sampled 100 contracts are false positives caused by the inaccuracy
    of the Mythril symbolic execution engine when handling storage accesses;
    the remaining 62 reported errors are benign, e.g., integer overflows that
    are later filtered by assertions and conditions.}.

Our experimental results sound the alarm on the prevalence of standard
violation errors, i.e., even the top smart contracts that manipulate critical
digital assets often do not fully conform to their corresponding standards.
Out of the 255 found errors, 7 errors correspond to severe vulnerabilities that
can be exploited by any user in the blockchain and the underlying digital
assets are immediately threatened; 47 errors correspond to owner backdoors
which are only exploitable by the contract owners; 20 errors correspond to
subtle constraint violations that are not immediately exploitable given the
current blockchain state but they may become exploitable as the blockchain
state changes; the remaining 181 errors are non-exploitable standard deviations
in customized features but they may cause undesirable experience if other users
or contracts interact with them. 

\subsection{Contributions}

This paper makes the following contributions:

\begin{itemize}
\item \textbf{{\systemname}:} This paper presents a novel security analysis
    tool, {\systemname}, which enables users to specify customized constraints
    and invariants to detect EVM smart contract errors that cause deviations
    between a contract implementation and the corresponding specification.

\item \textbf{Symbolic Execution for EVM Bytecode:} This paper presents a
    symbolic execution framework for EVM byte code. In particular, it presents
    novel techniques for efficiently handling EVM instructions that access the
    persistent state and the volatile memory.  

\item \textbf{Evaluation of {\systemname}: } This paper presents a systematical
    evaluation of {\systemname} on a benchmark set of 1089 smart contracts.
    {\systemname} found 255 errors in total and 237 out of the 255 errors are
    zero-day errors.  

\item \textbf{Standard Violation Alarm:}
    This paper is the first to identify the prevalence of standard violation
    errors in critical smart contracts that manipulate publicly traded digital
    assets.  The total market cap of these vulnerable zero-day smart contracts
    is more than nine hundred million dollars at the time of writing this
    paper\footnote{February 12, 2019}. This paper also presents analyses and
    case studies on these standard violation errors we found to guide future
    efforts to eliminate such errors in smart contracts. 
\end{itemize}

The remainder of this paper will be organized as follows.
Section~\ref{sec:example} presents a motivating example to illustrate
{\systemname}. Section~\ref{sec:design} presents the technical design of
{\systemname}.  Section~\ref{sec:impl} discusses the implementation of
{\systemname}.  We evaluate {\systemname} with experiments and analyze standard
violation errors we found with {\systemname} in Section~\ref{sec:results}.  We
finally discuss related work in Section~\ref{sec:related} and conclude in
Section~\ref{sec:conclusion}.

\section{Example}
\label{sec:example}

\begin{figure}[t!]
\begin{minted}{JavaScript}
contract BecToken {
  uint256 public totalSupply;
  mapping(address => uint256) public balanceOf;
  
  function transfer(address _to, uint256 _value) public returns (bool) {
    require(_to != address(0));
    require(_value > 0 && _value <= balanceOf[msg.sender]);
    require(_value + balanceOf[_to] > balanceOf[_to]);
    balanceOf[msg.sender] = balanceOf[msg.sender] - _value;
    balanceOf[_to] = balanceOf[_to] + _value;
    Transfer(msg.sender, _to, _value);
    return true;
  }

  function batchTransfer(address[] _receivers, uint256 _value) public whenNotPaused returns (bool) {
    uint cnt = _receivers.length;
    /* Integer overflow */
    uint256 amount = uint256(cnt) * _value;
    require(cnt > 0 && cnt <= 20);
    require(_value > 0 && balanceOf[msg.sender] >= amount);
    balanceOf[msg.sender] = balanceOf[msg.sender].sub(amount);
    for (uint i = 0; i < cnt; i++) {
        balanceOf[_receivers[i]] = balanceOf[_receivers[i]].add(_value);
        Transfer(msg.sender, _receivers[i], _value);
    }
    return true;
  }
}
\end{minted}
\caption{Simplified source code from BECToken.}
\label{code:BECToken}
\end{figure}

In this section, we present a motivating example of how {\systemname} detects an
invariant violation error in the smart contract of BECToken.  Figure
\ref{code:BECToken} presents the simplified source code of this example.
BECToken is an ERC-20 contract~\cite{ERC20} deployed on Ethereum blockchain.
ERC-20 is a technical standard that defines a set of contract interface
functions to implement a digital token asset. Specifically in an ERC-20
contract, the public property or the function \sourcecode{balanceOf()} should
return the amount of tokens that the given address owns (line 3 in
Figure~\ref{code:BECToken}); the function \sourcecode{transfer()} should
transfer the specified amount of tokens from the address of the message sender
 to the specified receiver address (lines 5-13); the public property
or the function \sourcecode{totalSupply()} should return the total amount of the
circulated tokens (line 2). An ERC-20 contract should also satisfy the
following invariant: the sum of token balances of all addresses should equal to
the total supply at any time. 

In Figure~\ref{code:BECToken}, the public \sourcecode{balanceOf} map stores the
balance of each address; the public \sourcecode{totalSupply} variable stores
the current total supply of the token. BECToken implements
\sourcecode{transfer()} accordingly with these two global variables.  Note that
these global variables will reside on the Ethereum blockchain permanently so
that any full node in Ethereum can access and verify the current BECToken state
including the balance of any address. Also note that \sourcecode{uint256} is a
special large integer type in Ethereum with 256 bits.

However, BECToken implements a customized batch transfer function called
{\sourcecode{batchTransfer()}}.  The intended behavior of
\sourcecode{batchTransfer()} is to transfer the specified amount of tokens from
the address of the transaction initiator to each address in a specified array.
Unfortunately, there is an integer overflow error at line 18, where the
statement calculates the total transferred amount. Specifically, if an attacker
creates a transaction and calls \sourcecode{batchTransfer()} with
\sourcecode{\_value} as $2^{255}$ and \sourcecode{\_receivers} containing two
addresses, \sourcecode{amount} would become zero after the overflow (recall
that amount has 256 bits).  This overflowed value in turn would enable the
attacker to bypass the security check at line 20. The consequence of this
attack is that the attacker could therefore send a large amount of tokens that he
or she does not own, effectively generating BECTokens from the air and violating
the ERC-20 standard.  Note that this vulnerability was exploited by an
anonymous attacker in 2018 April, who sent massive amount of generated coins to
crypto-exchanges for profit at the expense of other honest token holders. The
market cap of BECToken, which was two billion USD at the time of the attack,
evaporated in days~\cite{bec}.

\begin{figure}[t!]
\begin{minted}{Python}
sum = 0
for address in ADDRS:
  bal = C.balanceOf(address)
  check(sum + bal >= sum)
  sum += bal
check(sum == C.totalSupply())
\end{minted}
\caption{User specified invariant.}
\label{code:invariant}
\end{figure}

\noindent{\textbf{Specify Invariant:}} We next apply {\systemname} to BECToken
and describe how {\systemname} could detect this vulnerability. We first
provide {\systemname} the total supply invariant we want to check as
Figure~\ref{code:invariant}. In {\systemname}, users write invariants in a
syntax similar to Python. In Figure~\ref{code:invariant}, the \sourcecode{C} keyword
is a predefined variable as the handle of the being checked contract;
\sourcecode{ADDRS} is another predefined variable to represent the set of all
possible addresses.  The invariant in Figure~\ref{code:invariant} iterates over
all addresses, calls \sourcecode{balanceOf()} to retrieve the balance of each
address, and then checks whether the sum of these balances equals to the result
returned by \sourcecode{totalSupply()}. There is also a special check at line 4
to make sure that the computation of the sum does not cause integer overflow
errors. 

Note that the user does not need the access of the contract source code in
Figure~\ref{code:BECToken} to write this invariant. The user only needs to know
the contract Application Binary Interface (ABI), which is specified by smart
contract standards and is typically published by the developers as well.
In fact, Figure~\ref{code:BECToken} is a
general-purpose invariant that not just the BECToken contract but all ERC-20 contracts
should satisfy.

\noindent{\textbf{Generate Function Constraints:}} Because the invariant in
Figure~\ref{code:invariant} should be satisfied before and after every function
invocation, {\systemname} automatically generates a function constraint from
the invariant for each interface function of BECToken.  Figure~\ref{code:total}
presents the generated constraint for \sourcecode{batchTransfer()}. 
At lines 1-7, the generated constraint ensures that the invariant holds for the
initial state of the contract before the invocation of
\sourcecode{batchTransfer()}.  At lines 8-11, the constraint invokes
\sourcecode{batchTransfer()} with a symbolic array of receiving addresses and a
symbolic integer amount.  \sourcecode{caller = SymAddr()} at line 11 also makes
the transaction initiated by a symbolic address as well. Note that symbolic
values here mean that these values are external inputs and reflect the fact
that anyone can initiate a transaction to call \sourcecode{batchTransfer()}
with arbitrary parameters. At lines 12-18, the generated constraint checks the
total supply invariant again to check that the invariant is satisfied after the
\sourcecode{batchTransfer()} call.
Note that users can also write constraints like those in Figure~\ref{code:total}
directly.  Such capability is useful when the user wants to use {\systemname}
to check dedicated properties of one function rather than general invariants.
In our experiments, we utilized this capability of {\systemname} to check
function specific properties of our benchmark contracts.  See
Section~\ref{sec:results} for more details.

\begin{figure}[t!]
\begin{minted}{Python}
# Invariant as the pre-condition
sum = 0
for address in ADDRS:
  bal = C.balanceOf(address)
  assume(sum + bal >= sum)
  sum += bal
assume(sum == C.totalSupply())
# Invoke the checked function
raddrs = SymArray()
amount = SymInt()
C.batchTransfer(raddrs, amount, sender=SymAddr())
# Check invariant as the post-condition
sum = 0
for address in ADDRS:
  bal = C.balanceOf(address)
  check(sum + bal >= sum)
  sum += bal
check(sum == C.totalSupply())
\end{minted}
\caption{Constraints generated by {\systemname}.}
\label{code:total}
\end{figure}

\noindent{\textbf{Symbolic Execution:}} {\systemname} then runs symbolic
execution analysis on the contract EVM bytecode to check each of the generated
constraint. The symbolic execution engine of {\systemname} initializes program
values in the global states as fresh symbolic variables similar to input
parameters of \sourcecode{batchTransfer()}.  The global states include
variables like \sourcecode{totalSupply} and \sourcecode{balances} in lines 2
and 3 in Figure~\ref{code:BECToken} and blockchain states like the block height
and timestamp that might be queried by the contract program.  {\systemname}
then symbolically executes the generated constraint and the contract EVM
bytecode to maintain a map that maps each program value to a symbolic
expression. The symbolic expression represents the computation of deriving the
corresponding program value from initial global states and input parameters.
{\systemname} also maintains a symbolic path constraint, which is the
constraint that initial global states and input parameters need to satisfy in
order to exercise the current execution path. The symbolic execution engine
updates the path constraint when it processes conditional branches and require
statements.

\noindent{\textbf{Detect Error:}} When the symbolic execution engine processes
the check statements (e.g., lines 16 and 18 in Figure~\ref{code:total}),
{\systemname} uses an off-the-shelf SMT
solver~\cite{NiemetzPreinerBiere-JSAT15} to determine whether it is possible
find an assignment scheme of initial global states and input parameters such
that the check condition is false.  If so, {\systemname} successfully detects
an error and the solver will return the error triggering assignment scheme as
the error report.

In our example, when {\systemname} processes the check statement at line 18 in
Figure~\ref{code:total}, the solver returns that it is possible to violate this
condition by setting \sourcecode{\_value} to a large number.
Appendix~\ref{appendix:error-triggering} presents a concrete example that
{\systemname} generates to violate the total supply invariant. With the above
assignment, the security check \sourcecode{balances(msg.sender)>=amount} in
line 20 in Figure~\ref{code:BECToken} is bypassed and \sourcecode{msg.sender}
is able to send more tokens to receivers than it owns. After the execution of
\sourcecode{batchTransfer()}, the check condition at line 18 in
Figure~\ref{code:total} is violated. {\systemname} therefore successfully
detects and reports this error to the user.

\section{Design}
\label{sec:design}
We next formally present the design of the {\systemname} symbolic execution engine.

\subsection{Core Language}
\label{sec:design:lang}

\begin{figure}
    \input{figures/core-lang.tex}
    \caption{Core Language}
\label{fig:core}
\end{figure}

A smart contract transaction is a tuple $T = \langle P, A \rangle$, where $P$
is the combined bytecode program of the invoked contract function and $A$ is
an input stack that contains all transaction input parameters including the
sender account address.

Figure~\ref{fig:core} presents the syntax of a simplified stack based virtual
machine language that we will use to illustrate the symbolic execution engine
of {\systemname} in this section.  Similar to the Ethereum Virtual Machine
(EVM) and other standard virtual machine languages, a program in this language
is a map that maps integers (i.e., program counters) to instructions.  The
program executes with an execution stack.  Given an initial execution stack
that contains input transaction values and an initial state of the persistent
blockchain storage, the program execution starts with the program counter being
zero and ends when the execution reaches the \texttt{stop} instruction.  All
program values during the execution are integers and similar to EVM all
integers have a fixed width of 256 bits.  All computations during the execution
also happen with 256 bits, e.g., an equal comparison instruction will produce a
256 bit integer zero if the two operands are not equal and a 256 bit integer one if
the operands are equal.

The language has four execution stack manipulation instructions, \texttt{push}
for pushing an extra constant into the stack, \texttt{pop} for popping the top
value from the stack, \texttt{swap} for swapping the last two elements, and
\texttt{dup} for duplicating the top value of the stack.  The language also has
various arithmetic and comparison instructions such as \texttt{add} and
\texttt{eq}, each of which pops values from the execution stack as operands and push
the computation result back.  \texttt{jumpi} is the jump instruction which
checks the top value at the stack and conditionally jumps to a specified
program counter. 

Instructions \texttt{addrof} and \texttt{addrofmap} are used to compute the
address of a variable, an array, or a map.  Each of these instructions is
followed with either \texttt{sload} or \texttt{sstore}, which are instructions
for accessing the persistent state of the contract.  The persistent state of the
contract is a flat map that maps 256 bit integer addresses to 256 bit integer
values.  \texttt{addrof} or \texttt{addrofmap} fetches a global variable slot
index and an array index or a map key form the stack, computes a cryptographic hash,
and pushes the hash result to the stack as the address.  The following
\texttt{sload} or \texttt{sstore} then accesses the memory at the computed
address.  Note that for brevity, the language in Figure~\ref{fig:core} omits
many features of EVM such as the volatile memory, the ether account balance,
the gas system, and miscellaneous blockchain states. Our {\systemname}
implementation supports all these features. We will describe how {\systemname}
handles the volatile memory in Section~\ref{sec:design:storage} and the
remaining EVM features in Section~\ref{sec:impl}.

The \texttt{assume} and \texttt{check} instructions in Figure~\ref{fig:core}
represent the corresponding functions in the user specified constraints. We
include them in the language so that we can illustrate how {\systemname}
verifies the constraints.  Note that our current implementation {\systemname}
performs symbolic executions on both of the EVM bytecode of the checked
contract and the user specified constraints together. See
Section~\ref{sec:impl:constraint}.

\subsection{Semantics for Symbolic Execution}
\label{sec:design:semantic}

\begin{figure*}[t]
    \input{figures/opsemantics2.tex}
    \vspace{-0.15in}
    \caption{Small-Step Operational Semantics of Instructions for Symbolic Values}
\label{fig:sseman}
\end{figure*}

\noindent{\textbf{Execution Environment:}} The environment of the symbolic
execution is a tuple $\sigma = \langle \mathit{pc}, K, S, \psi \rangle$.
$\mathit{pc}$ is the current program counter value which could be an integer,
\texttt{nil} (indicating normal termination), or \texttt{err} (indicating
errors).  $K$ is a stack of symbolic expressions to represent the execution
stack.  The symbolic execution differs from the normal execution in that each
concrete program value in the normal execution is instead replaced by a
symbolic expression.  This symbolic expression tracks the sequences of
computations of generating the program value from initial state values and
input values. 

The symbolic array $S$ represents the persistent state. A symbolic
array has two operations \texttt{Store()}and \texttt{Load()}.
$\texttt{Store}(S, E_1, E_2)$ updates a symbolic array by storing the symbolic
expression $E_2$ at the symbolic index (expression) $E_1$ and returns the
updated array.  $\texttt{Load}(S, E_1)$ returns the symbolic expression that
corresponds to the last stored value at the symbolic index (expression) $E_1$.
The symbolic expression $\psi$ denotes the current path constraint.
The path constraint in the symbolic execution denotes the
constraint that the initial state values and the input values must satisfy to
make the current execution path feasible. See Figure~\ref{fig:core} for the syntax of 
symbolic variables, expressions, and arrays.

\noindent{\textbf{Small Step Semantics:}} Figure~\ref{fig:sseman} presents the rules of
the small step semantics for our symbolic execution. Each rule is of the form
$\sigma \lll I \rrr \sigma'$.  It denotes that given the original execution
state $\sigma$, after executing the instruction $I$ the new execution state can
be $\sigma'$. The rules for $\texttt{push}$, $\texttt{pop}$, $\texttt{swap}$,
$\texttt{dup}$, $\texttt{add}$, and $\texttt{eq}$ are self-explanatory. They
fetch operands from the current execution stack and update the stack with the
results accordingly.  

The rule for $\texttt{assume}$ updates the path constraint with the assumed
constraint.  There are two rules for $\texttt{jumpi}$, one for the true branch
and the other one for the false branch. Besides changing the program counter,
these rules also update the path constraints to indicate additional constraints
on the input and initial state values for exercising the corresponding
execution paths. The rule for $\texttt{check}$ is similar to $\texttt{jumpi}$
and $\texttt{assume}$. But if the checked constraint fails, the rule updates
the program counter to $\texttt{err}$.

The rules for $\texttt{addrof}$ and $\texttt{addrofmap}$ take two operands from
the stack and use a hash function $\mathrm{Hash}$ to compute the address for
the followed store instructions.  The rule for $\texttt{sstore}$ updates the
symbolic array with a $\texttt{Store}$ operation and the rule for
$\texttt{sload}$ pushes a fresh symbolic variable $x$ to the stack to represent
the loaded value and puts an additional constraint to ensure that $x$ equals to
the value that a $\mathrm{Load}$ operation on $S$ returns.  Note that
$\mathrm{Hash}$ denotes a special hash function that converts two symbolic
expressions (corresponding to the global variable index and the array index or
the map key) into a single 256 bit address. This faithfully matches the
behavior of the EVM bytecode generated by the compiler of
Solidity~\cite{solidity}, the most popular high-level language for developing
Ethereum smart contracts.  The compiler generates code to use
SHA3~\cite{wood2014ethereum} function to determine the final address of a
stored object.

\subsection{Main Algorithm}
\label{sec:design:algo}

\begin{algorithm}[t]
    \SetNlSty{}{}{}
\DontPrintSemicolon
\SetKw{To}{ to }
\SetKw{In}{ in }
\SetKw{Not}{ not }
\SetKw{Continue}{continue}
\SetKwInOut{Input}{Input}
\SetKwInOut{State}{Local State}
\SetKwFor{Event}{upon event}{do}{}
\let\oldnl\nl
\newcommand{\nonl}{\renewcommand{\nl}{\let\nl\oldnl}}
\Input{A list of $n$ transactions $T_1$, $T_2$, $\ldots$, $T_n$, where $T_i = \langle P_i, A_i\rangle$}
$\sigma_0 \leftarrow  \langle 0, A_1, s, \texttt{true}\rangle$, where $s$ is a free symbolic array\;
$\Gamma \leftarrow \{\langle \sigma_0, 1 \rangle\}$\;
\While{$\Gamma \neq \emptyset$}{
    Select and remove a state $\langle \langle pc, K, S, \psi\rangle, i\rangle$ from $\Gamma$\;
    $\Sigma' \leftarrow \{\sigma' | \langle pc, K, S, \psi \rangle \lll P_i(pc) \rrr \sigma\}$\;
    \For{$\sigma' = \langle pc', K', S', \psi' \rangle \in \Sigma'$} {
        \If {\Not $\mathrm{Satisfiable}(\psi')$} {
            \Continue
        }
        \If {$pc' = \texttt{err}$}{
            $\mathrm{GenerateReport}(\sigma')$\;
        }
        \ElseIf {$pc' \neq \texttt{nil}$} {
            $\Gamma \leftarrow \Gamma \cup \{\langle \sigma', i \rangle\}$\;
        }
        \ElseIf {$i \neq n$}{
            $\Gamma \leftarrow \Gamma \cup \{\langle \langle 0, A_{i+1}, S', \psi'\rangle, i+1 \rangle\}$\;
        }
    }
}

    \caption{Main loop}
\label{code:main}
\end{algorithm}

Figure~\ref{code:main} presents our main symbolic execution algorithm. The
input to the algorithm is a list of transactions, because a user specified
constraint can check the behavior of multiple transactions in sequence.
$\Gamma$ in Figure~\ref{code:main} maintains the set of active execution states
and we initialize $\Gamma$ with the input stack of the first transaction as the
execution stack and a fresh symbolic array as the persistent state at line 1.
Each iteration of the main loop at lines 3-14 selects one active state
$\sigma$ from $\Gamma$ and symbolically executes on $\sigma$ for one step with
the rules in Figure~\ref{fig:sseman}.  For each possible result state
$\sigma'$, it first invokes the underlying SMT solver to determine whether the
path constraint is satisfiable at line 7.  If not, then it is discarded as an
impossible state. Then it checks whether the program counter reaches
\texttt{err}, which indicates a constraint violation.  If so, it will generate
an error report. If not, it pushes the new state back
to $\Gamma$.

When selecting states in line 4 in Figure~\ref{code:main}, {\systemname}
uses the depth first strategy and always selects the newly-generated state.
Instead of bounding the number of loop iterations to make the symbolic execution in
Figure~\ref{code:main} computationally tractable, {\systemname} provides each transaction limited gas and the transaction
aborts if the gas is exhausted. The rationale is that most
errors can be triggered with a limited number of iterations and a small size of
array. 

\subsection{Storage Access Optimization}
\label{sec:design:storage}

\noindent{\textbf{Address Scheme:}} Handling the cryptographic hash functions in the
address computation in EVM is challenging.  If we encode SHA3 computations
directly into the symbolic expression, the resulted expressions will be too
complicated for any SMT solver to solve.  One naive approach is to give up and
return a fresh symbolic variable to represent the loaded value when facing
complicated symbolic addresses.  This approach is adopted by previous work like
Mythril~\cite{mythril}. It makes the analysis results inaccurate and may
introduce many false positives.  Another naive approach is to concretize
address values to avoid handling cryptographic computations symbolically. This
approach is adopted by other previous work~\cite{trailofbits-manticore}, but
such concretization may cause the symbolic expression size to grow
exponentially when analyzing contracts with multiple storage accesses.

{\systemname} instead replaces the address scheme in the $\texttt{addrof}$ and
$\texttt{addrofmap}$ with a customized function shown as
Figure~\ref{code:hash}. The intuition is to avoid collisions of different
global objects without using expensive cryptographic hash functions.  
{\systemname} expand
the address space of the persistent storage with additional bits to fit both of
the global slot index and the key value.  This guarantees no collision of
different storage objects and it is functionally equivalent to the original
contract behavior (as long as attackers cannot find collisions in SHA3 to
trigger object collisions in the original contract). More importantly, the
computation in Figure~\ref{code:hash} is much cheaper for an SMT solver to
handle.

In the current implementation, {\systemname} bounds the size of any dynamically
sized objects (i.e., arrays and maps) to three. This enables {\systemname} to
focus on exploring new states that more likely contain errors than less
productive states in iterative structures. Although the bound in theory may
cause the analysis to miss some errors, in practice we found no error requiring
more than three elements in dynamically sized objects to trigger in our
experiments.

This bound also enables {\systemname} to further reduce the number of bits used
in the computation in Figure~\ref{code:hash}. The bound limits the maximum
index of the stores for arrays ($E_2$ in Figure~\ref{code:hash}).  For common
map cases, the map keys are Ethereum account addresses. {\systemname} collects
all possible account addresses into an address pool (see
Section~\ref{sec:impl:pool}) and normalize them into bounded integer indexes as
well. With these two optimizations, {\systemname} actually does not need to
shift 256 bits to avoid collisions. In our current implementation,
{\systemname} keeps the computed addresses as 256 bit vectors to simplify the
resulting symbolic expressions.

\begin{algorithm}[t]
    \SetNlSty{}{}{}
\DontPrintSemicolon
\SetKw{To}{ to }
\SetKw{In}{ in }
\SetKw{Not}{ not }
\SetKw{Continue}{continue}
\SetKwInOut{Input}{Input}
\SetKwInOut{Output}{Output}
\let\oldnl\nl
\newcommand{\nonl}{\renewcommand{\nl}{\let\nl\oldnl}}
\underline{function} Hash$(\mathit{E_1}, \mathit{E_2}):$\\
\Input{The global slot index $\mathit{E_1}$ and the index for the store $\mathit{E_2}$}
\Output{The hashed address $\mathit{E'}$}
$E' \leftarrow \mathrm{BitExtend}(E_1,256)$\;
$E' \leftarrow \mathrm{ShiftLeft}(E', 256) + E_2$\;

    \caption{Customized address scheme function.}
\label{code:hash}
\end{algorithm}


Note that {\systemname} distinguishes maps from other dynamic objects in EVM
bytecode with the following heuristics. In EVM, the maps are not stored
sequentially and the map keys participate the cryptographic computation for
determining the final addresses of mapped values.  In contrast, other objects
like array are stored sequentially. {\systemname} therefore statically analyzes
EVM instruction sequences before each load and store to determine the data type
of the accessed object.

\noindent{\textbf{Volatile Memory:}} 
Besides global persistent state $S$ described in this section, EVM also
contains volatile memory to store temporary data which will be wiped after each
transaction execution finishes.  {\systemname} models memory using a symbolic
array similar to the global persistent state. Note that unlike the persistent
state, the memory does not use complicated cryptographic hash functions 
for computing
addresses so that the address scheme is continuous and simpler.

However, one challenge of handling the memory is that unlike the rest of EVM
programs which use 256 bit integer values, the memory maps 256 bit address to 8
bit values.  Therefore storing a 256 bit value from stack to the memory
requires the system to break the value into 32 bytes and store each of them
separately.  Naively processing these data transfer instructions will cause the
symbolic execution engine to generate numerous redundant bit
operations for symbolic expressions if the checked constraints contain
intensive memory operations.  Appendix~\ref{appendix:memory} presents such an
example.  Such redundant operations will slow down the underlying SMT solver
dramatically.  To address this challenge, {\systemname} maintains a cache to
optimize volatile memory access during symbolic executions. {\systemname}
detects instruction patterns like Appendix~\ref{appendix:memory} which store
and load 256 bit values into and from the memory. When the program stores a
symbolic 256 bit value into the memory, the cache records the address and the
256 bit symbolic value. When the program loads a symbolic 256 bit value from
the memory, {\systemname} queries the SMT solver to determine whether the
loaded address corresponds to only one possible cached address from the cache.
If so, {\systemname} returns the corresponding symbolic value from cache
directly without generating redundant bit extraction and concatenation
operations. 


\noindent{\textbf{Results:}} Our experimental results show that our storage
access optimizations increase the efficiency of the system dramatically. In our
experiments, the combination of the address scheme optimization and the
volatile memory optimization enable {\systemname} to find out 67 more errors
and/or to generate much less false positives than other alternative approaches.
See Section~\ref{sec:results}. 

\section{Implementation}\label{sec:impl}

We built {\systemname} on top of Manticore~\cite{trailofbits-manticore}, an
open source symbolic execution framework. The original Manticore does not 
fully support EVM instructions, does not support user defined constraints,
and handles storage access addressing inefficiently with concretization.
Besides the algorithms described in Section~\ref{sec:design}, {\systemname} is
extended to support all EVM instructions described in Ethereum yellow
paper~\cite{wood2014ethereum}.  {\systemname} uses
boolector~\cite{NiemetzPreinerBiere-JSAT15} as its underlying SMT solver to
check satisfiability of the generated symbolic constraints during symbolic
executions.

\subsection{EVM Bytecode}\label{sec:impl:EVM}

\noindent\textbf{Indirect Jump:} One difference between EVM and the core
language in Section~\ref{sec:design} is that the jump instructions in EVM are
indirect, i.e., an EVM jump instruction fetches the jump target address from
the execution stack.  To handle such jump instructions, {\systemname} checks
whether the destination from the stack is symbolic. If so, {\systemname} gets
all possible values from the SMT solver and forks the current execution state
for each jump destination.  In practice, the current implementation of the Solidity
compiler guarantees that the jump destination is always constant.

\noindent\textbf{Miscellaneous Blockchain State:} Global blockchain states such
as timestamp and block height are initialized as fresh symbolic variables.  This
is consistent with the fact that a transaction may occur at any time and any
state in Ethereum.  {\systemname} updates these state values accordingly when
new transactions are generated. Each transaction is assigned with enough gas to
execute and {\systemname} leverages the gas limit system so that the system will stop
after certain iterations.

\subsection{User Constraints}\label{sec:impl:constraint}

User defined policies consist of 1) contract function calls, 2) assumptions,
and 3) safety constraints.  {\systemname} allows users to write Python scripts
to interact with a contract.  Calling a contract function is similar to normal
function calls, and the return value is deserialized from transaction return
data.  {\systemname} then translates those function calls into transactions and
symbolically executes them in sequential order.  A user can use
\sourcecode{assume(expr)} to define the preconditions and assumptions of
transaction arguments and initial states. The user can also use
\sourcecode{check(expr)} to define safety constraints that transaction
executions must satisfy. {\systemname} handles assumptions and constraints in
the same way as \sourcecode{assume} and \sourcecode{check} instructions
described in Section~\ref{sec:design}.

\subsection{Address Pool}
\label{sec:impl:pool}

In the Python script of a user defined constraint, the \sourcecode{ADDRS} keyword
represents an address pool that the script can interact with (as shown in
Figure~\ref{code:total}).  When {\systemname} starts, it first populates the
address pool with a fixed number of random addresses.  {\systemname} then
statically analyzes the EVM bytecode and detects additional addresses using
the following heuristic.  Different from other integer types, addresses in EVM
are 160 bits and the usage of each address is accompanied by a bit operation
to extract the last 160 bits from a 256 bit integer from the stack.
{\systemname} collects all detected concrete addresses into the address pool as
well.


\subsection{Inter-Contract Call}
Smart contracts in Ethereum can call functions defined in other contracts.
There are two usage scenarios of such inter-contract calls. The first one is to
reuse code in library contracts. The second one is to send transactions to
other smart contracts. To handle an inter-contract call, {\systemname} first
assumes the call belongs to the first scenario.  {\systemname} fetches and
executes the bytecode from the Ethereum blockchain. To keep the search space
included a reasonable size, {\systemname} takes advantage of storage access
optimization. If the callee address is a symbolic value and from the global
storage, {\systemname} fetches the current stored value from the Ethereum blockchain
instead of solving all possible values. If this address concretization fails,
then {\systemname} conservatively assumes that the call belongs to the second
scenario and {\systemname} will use a fresh symbolic value to represent the
returned value.

\noindent \textbf{Reentrance Errors:} One limitation of relying on user
speicifed constraints is that the user constraints may not specify side effect
that an external contract call could trigger. The constraints therefore may
miss reentrance errors. To this end, {\systemname} implements the ``no write
after external call'' rule similar to previous work~\cite{securify} and can
detect reentrance errors with this rule.  Specifically, the symbolic execution
engine in {\systemname} will generate a reentrance error warning if there is an
execution path that the contract first 1) calls a non-library external function
and then 2) modifies the persistent state. 


\section{Evaluation}\label{sec:results}

We next evaluate {\systemname} with a set of 1089 real world smart contracts.
The goal this evaluation is to answer the following questions:
\begin{enumerate}
    \item How effective is {\systemname} in finding standard violation errors in Ethereum smart contracts?
    \item What kinds of errors {\systemname} find and how many of them are zero-day errors?
    \item How does {\systemname} compare with previous smart contract analysis tools?
    \item How much improvement do the storage access optimizations in
        Section~\ref{sec:design:storage} have on the performance of
        {\systemname}?
\end{enumerate}


\subsection{Methodology}\label{sec:method}

\noindent{\textbf{ERC-20 and ERC-721:}} ERC-20~\cite{ERC20} and
ERC-721~\cite{ERC721} are two important smart contract standards, which
define the contract interface and specification for implementing fungible and
non-fungible digital assets respectively.  Those two standards are
well-established and followed by many smart contracts in Ethereum. Many of the
underlying digital assets of these contracts are publicly traded in
crypto-exchanges.  Given the importance and the popularity of ERC-20 and
ERC-721, we therefore focus our evaluation on contracts following these two
standards. 

\noindent{\textbf{Collect Benchmark Contracts:}} 
To collect a representative data set, we downloaded all ERC-20 and ERC-721
token contracts listed by Etherscan~\cite{etherscan} on November 29, 2018 and
obtained in total 1089 contracts. We limit our study on these top contracts in
Etherscan because 1) we want to focus our analysis on important constracts that
manipulate critical digital assets and 2) many of the remaining ERC-20 and
ERC-721 constracts in the blockchain are experimental or duplicate.  Besides
the contract bytecode, we downloaded the application binary interface (ABI) of
each contract as well as the contract source code.  The ABI lists all public
functions defined by a contract, which enables {\systemname} to call contract
functions and build symbolic transaction inputs.  We downloaded the source code
from Etherscan in order to analyze and classify the vulnerabilities.  Note that
this is for our manual analysis and {\systemname} does not require the source
code to check a contract. 

\noindent{\textbf{Specify Constraints:}} We developed one invariant and two
constraints based on the semantic of ERC-20 interfaces~\cite{ERC20}:
\begin{itemize}
\item \textbf{Total Supply:} The total supply invariant requires that no matter
    which function is executed, the sum of account balances equals the total
    supply of the token.
\item \textbf{Transfer:} The \sourcecode{transfer()} function succeeds if and only if sufficient tokens are supplied and the balances of the receiver and 
sender should be updated accordingly.
\item \textbf{Approval:} The \sourcecode{transferFrom()} function succeeds if and only if the token owner authorizes the message sender to do so and has a sufficient balance.
Both account balances and allowance of the sender should be updated if 
the function \sourcecode{transferFrom()} returns true.
\end{itemize}

Similarly, we developed one constraint based on the semantic of ERC-721
interfaces~\cite{ERC721}.  The \sourcecode{transferFrom()} function succeeds if
and only if the token owner authorizes the message sender to do so.  A detailed
explanation on how a user checks the invariants and the constraints using
{\systemname} will be provided in Appendix~\ref{sec:constraints}.

\noindent{\textbf{Apply {\systemname}:}} We apply {\systemname} to downloaded
contracts to check the developed constraints. In our experiments, we turned off
the reentrance error detector in {\systemname} to focus on constraint violation
errors.  All experiments are performed on a slurm cluster with two eight-core
Intel Xeon E5-2680 processors and 96GB rams on each node. Specifically,
{\systemname} allocates 20 minutes and 3GB memory for each contract. For each
error, {\systemname} reports the function name, the violated constraint, and a
concrete transaction trace that triggers the violation.

Note that in order to evaluate the effectiveness our optimizations in
{\systemname} for handling storage load and store instructions (see
Section~\ref{sec:design:storage}),  we implemented a baseline version of
{\systemname} that does not enable the optimizations.  For the persistent
state, the baseline implementation attempts to concretize the address values
whenever it encounters a load and store instruction paired with
cryptographic functions.  For the volatile memory, the baseline implementation naively
process every load and store instruction without the cache.  We apply this
baseline version of {\systemname} to all of our benchmark contracts and compare
the results with the normal version of {\systemname}.

\noindent{\textbf{Apply Mythril:}} We compare our results with
Mythril~\cite{mythril}, the state-of-the-art open source security tool
developed by Ethereum community.  Mythril is able to detect integer overflow
and reentrance errors. We download Mythril 0.20.0 which is the latest
version as available on January 30, 2019.  We run Mythril with the default
configuration to analyze all collected smart contracts.

\begin{figure*}[ht]
\begin{center}
\begin{tabular}{ |c|c|c|c|c|c|c|c| } 
\hline
\textbf{Policy} & \textbf{Severe} & \textbf{Backdoor} & \textbf{Attackable} & \textbf{Deviation} & \textbf{Missing} & \textbf{False Positives} & \textbf{Execution Time(s)}\\ 
\hline
Total & 2/1 & 47/33 & 20/20 & 87/87 & 19  & 3 & 187.47 \\
\hline
Approve (ERC-20) & 4/2 & 0/0 & 0/0 & 15/15 & 15 & 0 & 372.33 \\
\hline
Transfer & 0/0 & 0/0 & 0/0 & 95/95 & 5 & 2 & 119.02 \\
\hline
ERC-20 Total& 6/3 & 47/33 & 20/20 & 181/181 & / & 3 & 185.71 \\
\hline
\hline
Approve (ERC-721) & 1/0 & 0/0 & 0/0 & 0/0 & 116 & 0 & 944.94 \\ 
\hline
\end{tabular}
\end{center}
\vspace{-0.10in}
\caption{Evaluation Summary}
\label{fig:eva}
\end{figure*}

\noindent{\textbf{Analyze Detected Errors:}} 
We manually analyze each reported error 1) to determine whether it is a true
positive, benign error, or false positive 2) to classify the error based on who
can trigger the error and the severity of the error and 3) to check whether the
error is reported before and contact the contract owner and relevant stake
holders about the issues we found. Note that we classify a report as a benign
error if the error is not exploitable and is intended in the context. For
example, an integer overflow that is later filtered by a check or an assertion
before introducing any side effect is counted as a benign error.

{\systemname} reports 255 standard violation errors and 197 vulnerable
contracts.  We classify the 255 errors into four categories:
\begin{itemize}
    \item \textbf{Severe:} Severe errors can be exploited by anyone and may
        lead to financial loss of contract participants.
    \item \textbf{Backdoor:} Backdoor errors provide the contract owner
        exploitable privileges that are not consistent with the standard.
    \item \textbf{Attackable:} Attackable errors are theoretically exploitable,
        but the attack has to be performed at specific time period or an
        attacker may need to acquire a large amount of digital assets.
    \item \textbf{Deviation:} Some contracts implement their own features that
        are not consistent with the standards. Although we believe these errors
        are not exploitable, they may cause financial loss to a user who
        interacts with these contracts, if the user assumes the contracts
        conform to the standards.
\end{itemize}

Mythril reports plenty of issues. Even for high severity issues, Mythril
reports 1115 errors from 513 ERC-20 contracts and 595 errors from 217 ERC-721
contracts, which makes it impossible to analyze them all manually.  We
therefore sample 100 ERC-20 contracts and analyze the errors reported by
Mythril on the sampled contracts. A detailed comparison can be found in
Section~\ref{comp-mythril}.

\subsection{Results}\label{sec:results:results}



We evaluate {\systemname} for all 1089 smart contracts for policies described
in Appendix \ref{sec:constraints}.  Figure \ref{fig:eva} summarizes our
experimental results.  Each row presents the results of the corresponding
invariant or constraint.  Note that the fifth row summarizes all errors from
ERC-20 contracts.
For columns 2-5, each column represents the results for one type of errors,
severe errors, owner backdoors, attackable errors, and standard deviations.
Each cell presents two numbers. The first one is the number of found errors and
the second one is the number of zero-day errors that are not reported before.
The sixth column shows the number of contracts that do not implement all
required functions to perform the analysis. These contracts are not conforming
to the standards to implement all necessary functions.  And the seventh column
represents the number of false positives reported by {\systemname}.  Note that
{\systemname} does not report any benign error in our experiments.  The last
column presents the average execution time of {\systemname} to report an error. 

\noindent{\textbf{Severe:}} {\systemname} found 7 errors that are classified 
as "Severe". There are three types of severe errors:
\begin{enumerate}
    \item Contracts implement the \sourcecode{approve()} function
     without providing the functionality that allows 
    a user to revoke the approval, which may lead to token loss if a user sets the wrong approval.
    \item Contracts allows a user to transfer others token without approval. 
    \item The \sourcecode{totalSupply} variable is defined multiple times in different 
    parent contracts and transaction functions refer to inconsistent 
    \sourcecode{totalSupply} instances.
\end{enumerate}

\noindent{\textbf{Backdoor:}} 47 out of 255 errors are classified as owner
backdoors. All of those errors provide the admin privileges to modify balances
of other accounts that are not allowed in the ERC-20 standard.

\noindent{\textbf{Attackable:}} {\systemname} detects 20 theoretically
attackable errors that the possibilities to exploit such contracts are rare.
However, reusing the source code of such contracts may increase the risk of
being attacked.

\noindent{\textbf{Deviation:}} Some contracts implement their customized
features and deviate from the ERC-20 standard. Such deviations might be intended
and are not exploitable on their own, but such contracts might be misused if
the user assumes the contract follows the standard. 
We classified 181 errors as standard deviations. 
They can be further classified into two types. 
Firstly, a contract may implement its own logic beside ERC-20 interfaces such as
locked token or transaction fees.  
Secondly, many contracts do not implement ERC-20 interfaces as specified. For
example, some contracts implement the \sourcecode{transfer()} function without any return
value. Based on the EVM specification, the function returns 0 if no return value is
given which means the \sourcecode{transfer()} function returns false even if the
transaction succeeds. Such a contract is vulnerable if it interacts with other
contracts, and they use the return value to check if a transaction
succeeds.




\noindent{\textbf{Baseline Algorithm:}} The baseline algorithm analyzes 6335
functions without any exceptions and reports in total 188 errors, which is
roughly twenty percent less than {\systemname}. We further examine that all
errors reported by the baseline algorithm are covered by {\systemname}. The
baseline algorithm fails to report many errors that {\systemname} does report
because it runs out of the time without the optimizations for handling the
persistent state and volatile memory.

Our results show that {\systemname} is effective in finding standard violation
errors.  {\systemname} flags 255 from 197 vulnerable tokens and the tokens have
a market capitalization of 2 billion US dollars.  237 out of the 255 errors are
new errors that are not reported before. Our results also show that the
symbolic execution engine of {\systemname} is fast due to our optimizations for
handling storage address schemes.  We observe that 78\% of functions can be
verified within 20 minutes by {\systemname}. Without the optimization, only
68\% of functions can be verified within 20 minutes. 

\subsection{Comparison With Mythril}\label{comp-mythril}


We run Mythril on all benchmark contracts. Due to the high volume of reports
generated by Mythril, we randomly sampled 100 smart contracts to manually
compare the results of Mythril with {\systemname}.  
Out of the 127
issues reported by Mythril, there is only one true positive, which corresponds
to a severe vulnerability that is also reported by {\systemname}.  The remaining 126
reported issues are either benign errors or false positives.

Mythril generates 64 false positives for the 100 contracts, because Mythril fails to
handle load instructions to persistent state with symbolic addresses.  In such
cases, Mythril always returns a new unbounded symbolic value. If this unbounded
value is further used for arithmetic operations, Mythril will very likely
generate a false report. This again highlights the importance of handling
persistent state accurately for symbolic execution of EVM bytecode.
Mythril also generates 62 benign errors for the 100 contracts. 
The most common cases correspond to integer overflows that are later filtered
by assertions or if statements before introducing any side effect.  For those
errors not reported by Mythril, they contain logical errors that cannot be
detected based on low level EVM semantics. Figure~\ref{code:vezt} presents one
example of such logical errors, which we will describe in details in
Section~\ref{sec:results:case}.


\begin{figure}[t]
\begin{minted}{JavaScript}
function transferFrom(address from, address to, uint value) returns (bool success) {
  if (frozenAccount[msg.sender]) return false;
  if(balances[from] < value) return false;
  if( allowed[from][msg.sender] >= value ) return false;
  if(balances[to] + value < balances[to]) return false;
  balances[from] -= value;
  allowed[from][msg.sender] -= value;
  balances[to] += value;
  Transfer(from, to, value);
  return true;
}
\end{minted}
\vspace{-0.15in}
\caption{The RemiCoin contract code snippet.}
\label{code:vezt}
\end{figure}



Our results show that {\systemname} is more accurate and effective than
Mythril. {\systemname} is able to find significantly more errors with much less
false positives than Mythril in our experiments.  This is because {\systemname}
has a more accurate symbolic execution engine and {\systemname} detects high
level standard violations instead of low level EVM errors, many of which could
be benign. 

\subsection{Case Studies}\label{sec:results:case}

We next present one representative case for each type of standard violation
errors that {\systemname} found. We also present a case where {\systemname}
generates a false positive due to the imprecision of the supplied constraints. 

\noindent\textbf{Severe:} Figure \ref{code:vezt} shows the
\sourcecode{transferFrom()} function from RemiCoin \footnote{contract address:
    0x7dc4f41294697a7903c4027f6ac528c5d14cd7eb} found by approval policy.
There is a logic error at line 4 that the conditional statement returns false
if the allowance is greater or equal than \sourcecode{value}.  The correct
behavior is to return false if the allowance is smaller. This error 
allows an attacker to transfer one account's tokens to the other without proper
approval. Note that {\systemname} successfully detected this error, but Mythril
cannot.  This is because this logic error is not associated with any other
low-level runtime error.

\begin{figure}[t]
\begin{minted}{JavaScript}
function mint(address _holder, uint _value) external {
  require(msg.sender == ico);
  require(_value != 0);
  require(totalSupply + _value <= TOKEN_LIMIT);
  balances[_holder] += _value;
  totalSupply += _value;
  Transfer(0x0, _holder, _value);
}
\end{minted}
\vspace{-0.15in}
\caption{The ATL contract code snippet.}
\label{code:ATL}
\end{figure}

\noindent\textbf{Backdoor:} To allocate tokens on the fly, many tokens
implement a mint function that allows the contract admin to issues new tokens to
others.  However, missing security checks may leave backdoors that allow the
contract admin to modify balances of other accounts to an arbitrary number.
Figure \ref{code:ATL} shows the mint function from ATL \footnote{contract
address: 0x78b7fada55a64dd895d8c8c35779dd8b67fa8a05}. The \sourcecode{mint()} function allows
the contract owner to allocate tokens to an address \sourcecode{\_holder}. The
function first verifies that the transaction is from contract owner (line 2),
and the value is not zero (line 3). The contract sets a token limit that the
total supply of this token should not be more than \sourcecode{TOKEN\_LIMIT}
(line 4). The function then updates the balance of \sourcecode{\_holder} and
\sourcecode{totalSupply}.

{\systemname} found two issues from this function.  Suppose the owner calls
\sourcecode{mint()} with a large \sourcecode{\_value}.  First, the guard that
token's total supply should less than \sourcecode{TOKEN\_LIMIT} can be bypassed
by an overflow attack.  This allows the owner to allocate more tokens than the
contract allows. Second, \sourcecode{\_holder}'s balance can be overflowed and
can be set to any number at line 5. This effectively allows  the owner to
change a user's balance to an arbitrary number (e.g., to confiscate tokens from
a user).

\begin{figure}[t]
\begin{minted}{JavaScript}
function migrate(uint256 amount) {
  Token(previousToken).transferFrom(msg.sender, this, amount);
  pendingMigrations[msg.sender].dateTimeCreated = now;
  pendingMigrations[msg.sender].amount = amount;
}
function claimMigrate() {
  balances[msg.sender] += pendingMigrations[msg.sender].amount;
  totalSupply += pendingMigrations[msg.sender].amount;
  delete pendingMigrations[msg.sender];
}
\end{minted}
\vspace{-0.15in}
\caption{The simplified RexToken code snippet.}
\label{code:rex}
\end{figure}

\noindent\textbf{Attackable:} 
Figure \ref{code:rex} shows the \sourcecode{migrate()} and
\sourcecode{claimMigrate()} function from RexToken\footnote{contract address:
    0xf05a9382A4C3F29E2784502754293D88b835109C}. Those two functions allow
RexToken users to migrate their tokens stored in previous contracts to this
contract. While executing the \sourcecode{claimMigrate() function}, it
increases the balance of the sender and the total supply of the token
respectively without any overflow check at lines 7 and 8. {\systemname} marks
this contract as vulnerable because the total supply invariant can be violated
if the sender has large amount of token in previous contract.  Although this is
theoretically exploitable, we investigated these two contracts on Etherscan
main chain and such exploitations are not possible with the current contract
state.

\begin{figure}[t]
\begin{minted}{JavaScript}
function transfer (address _to, uint256 _value) returns (bool success) {
  uint256 fee = fee();
  if (accounts [msg.sender] < _value) return false;
  if (_value > fee && msg.sender != _to) {
    accounts [msg.sender] = safeSub (accounts [msg.sender], _value);
    accounts [_to] = safeAdd (accounts [_to], safeSub(_value, fee));
    accounts [fund] = safeAdd (accounts [fund], fee);
    tokensCount = safeSub (tokensCount, burnFee);
    Transfer (msg.sender, fund, fee);
    Transfer (msg.sender, _to, safeSub(_value, fee));
  }
  return true;
}
\end{minted}
\vspace{-0.15in}
\caption{The ParagonCoinToken code snippet.}
\label{code:paragon}
\end{figure}

\noindent\textbf{Deviation:} 
Figure \ref{code:paragon} shows the simplified \sourcecode{transfer()} function
from ParagonCoinToken\footnote{contract address:
0x7728dFEF5aBd468669EB7f9b48A7f70a501eD29D}. Unlike other ERC-20 tokens,
ParagonCoinToken computes a transaction fee each time a user transfer tokens to
the other (line 6-7), which breaks the transfer policy that all tokens should
be transferred to receiver's account. We believe this a deviation from ERC-20
standard. Note that a user can change the constraint to avoid treating this
constraint as vulnerable if desired.

\noindent\textbf{False Positive:}
Figure~\ref{code:fp} shows the \sourcecode{balanceOf()} function from
PRASMToken\footnote{contract address:
    0x1a66e09f7dccc10eae46e27cfa6b8d44a50df1e7}. {\systemname} reports that
this contract may break the specified constraints if an integer overflow occurs
when computing the balance at line 8 in Figure~\ref{code:fp}. This is
unfortunately a false positive because the total supply of this contract is
fixed as a constant after being initialized during the contract construction at line
4. Therefore, the sum of \sourcecode{balances[\_holder]} and
\sourcecode{lockupBalance} at line 8 will never overflow. {\systemname}
generates this false positive because the general purpose constraints we used do
not capture the precondition of the total supply. In fact, a more precise set
of constraints will enable {\systemname} to eliminate all three false positives
in our experiments.

\begin{figure}[t]
\begin{minted}{JavaScript}
constructor() public {
    ...
    initialSupply = 4000000000;
    totalSupply_ = initialSupply * 10 ** uint(decimals);
    balances[owner] = totalSupply_;
}
function balanceOf(address _holder) public view returns (uint256 balance) {
  return balances[_holder] + lockupInfo[_holder].lockupBalance;
}
\end{minted}
\vspace{-0.15in}
\caption{The PRASMToken code snippet.}
\label{code:fp}
\end{figure}

\subsection{Summary}

Our experimental results show that standard violation errors are prevalent in
Ethereum smart contracts. Over one sixth of evaluated ERC-20 and ERC-721
contracts contain one or more standard violation errors.  We found that for
many contracts that are supposed to follow a standard, their implementations in
the end deviate from the intended standard behavior.  This surprising result
calls for more efforts from the community to validate the consistency between
the contract standards and the actual implementations. We also have the
following important findings in our experiments.

\noindent \textbf{Severe Logic Errors:} Logic errors like
Figure~\ref{code:vezt} in a smart contract can cause severe vulnerabilities
that immediately threaten the security of the underlying digital assets
represented by the contract. This poses a challenge to standard analysis tools
that focus on specific types of low-level runtime errors.  {\systemname} can
successfully detect these logic errors, while previous tools like
Mythril~\cite{mythril} may miss them. This is because in these logic error
cases, the implementation simply does not match the intended behavior and no
low-level runtime error is triggered.

\noindent \textbf{Neglected Owner Backdoors and Corner Cases:} {\systemname}
finds a large amount of owner backdoor errors. This shows that software errors
in the owner privileged functions are often neglected by the community.  This
is undesirable because, in ERC-20 and ERC-721 contracts, the owner corresponds
the issuer of the digital assets. Therefore, the analogies of these owner
backdoors are intentional or unintentional hidden text in a legal contract for
financial assets. 

{\systemname} also finds a large amount of theoretically attackable contracts.
This shows that the smart contract developer has a tendency to ignore corner
cases.  For example, one vulnerable contract casts a balance value of 256-bit
integers directly down to 128-bit integers without proper checks, assuming the balance
will always be less than 128 bits. Consider the fact that smart contract code is
often copied and reused by multiple contracts, such code may eventually be
exploited when relevant conditions are satisfied.

\noindent \textbf{Error-prone Customized Features:} We also found that the
implementations of customized features in smart contracts are often vulnerable
or inconsistent with the corresponding standards. We believe the smart contract
developers should proceed with extreme caution when designing and implementing
customized features.

\noindent \textbf{Developer Irresponsiveness:} Last but not least, we found
that smart contract developers/owners are not very responsive for error
reports. One explanation is that developers and owners do not have the
capability to quickly fix a reported issue. Ethereum does not allow any update
to a deployed smart contract.  The rationale is that the smart contract
corresponds to the law that all participants should follow and that no one
should be able to change the law. The only way to fix a vulnerable contract is
to deploy a new contract and to persuade every participant to accept the new
contract instead of the old one. Another explanation is that as the issuers of
publicly traded digital assets, developers and owners may be economically
incentivized to hide a smart contract error if the exploitation chance is low.

\section{Related Work}\label{sec:related}

\noindent\textbf{Smart Contract Security:}
Researchers recently have proposed several automated program analysis
techniques to help detect security errors and vulnerabilities in the smart
contracts.  Oyente~\cite{Luu:2016:MSC:2976749.2978309} implements a symbolic
execution tool that detects vulnerability patterns of transaction-ordering
dependency attacks, timestamp attacks, reentrance attacks, and mishandled
exception attacks.  Mythril~\cite{mythril} and
Manticore~\cite{trailofbits-manticore} are open source symbolic execution
frameworks for detecting integer overflows and reentrance errors.  Kolluri et
al. use symbolic execution techniques to detect races of smart contract
transactions~\cite{ethracer}.  Maian~\cite{DBLP:journals/corr/abs-1802-06038}
analyzes traces with symbolic execution to detect vulnerable contracts that
handles ether transfers.  A more recent work, Zeus~\cite{kalra2018zeus},
        converts solidity source code into a customized low level virtual
        machine language that is compatible with LLVM~\cite{llvm} and then uses
        existing taint analysis and symbolic execution tools in LLVM to detect
        common vulnerability patterns and fairness issues.  {\systemname}
        differs from these symbolic execution tools in that instead of focusing
        on specific patterns of runtime errors, {\systemname} validates the
        consistency between a contract implementation and the corresponding
        standard.  {\systemname} also works with optimizations on handling EVM
        storage access instructions for more efficient symbolic execution of
        EVM byte code directly.  Together, these differences enable
{\systemname} to find significant more errors (including logic errors) with
much less false positives in our experiments. 



ContractFuzzer~\cite{contractfuzzer} proposes a fuzzing framework for detecting
seven kinds of vulnerabilities from Ethereum smart contracts.  Comparing to
{\systemname}, fuzzing techniques may miss many errors (i.e., introduce false
        negatives) because they only exercise a limited number of execution
paths.  Securify~\cite{securify} provides a domain-specific static analysis
framework for smart contracts, that translates EVM bytecode into Datalog and
use off-the-shelf verifier to check predefined security properties like that no
storage write should occur after function calls and that all ether transfers
should be protected by conditions.  TEEther~\cite{teether} uses static analysis
to detect control flow paths that lead to attackable instructions that
attackers can exploit to steal ether. Comparing to {\systemname}, these static
analysis techniques target specific patterns of errors, while {\systemname} can
detect any deviation between an implementation and the corresponding standard.

KEVM~\cite{kevm} formally defines the EVM semantics in $\mathbb{K}$ and
EVM*~\cite{evmstar} translates EVM bytecode to F*.  Building EVM semantics
allows a user of $\mathbb{K}$ or F* to build their own policy for further
verification. VeriSolid~\cite{verisolid} is a model-based approach that allows
a user to specify a transition model as well as security rules.  Similarly,
  Vandal~\cite{vandal}, a static analysis framework, translates EVM bytecode to
  logic relations and detects vulnerabilities described in Souffl\'e language.
  Unfortunately, these verification tools either cannot handle full EVM byte
  code or require significant human interventions during the verification
  process.


\noindent\textbf{Contract Development Tools:} Libraries such as
openzeppelin-solidity~\cite{openzeppelin} provide safe arithmetic operations
which avoid integer overflow and underflow attacks. It also provides a
reference implementation of standard protocols.
However, using such libraries is not mandatory.  Ironically, BecToken included
SafeInt library in its contract code, but it does not use the library for all
arithmetic operations. Moreover, such libraries can not prevent contracts
from logic errors such as missing authorization.

Breindenbach et al. propose the hydra framework, automates the process of
discovering bugs and distributing bounties~\cite{enterthehydra}, which
incentivizes the bug disclosure. Erays~\cite{erays} is a reverse engineering
tool that decompiles EVM bytecode into high-level solidity-like pseudocode. A
user can further analyzes the security issues inside the contract without
accessing to the contract source code.  Wang et al. propose a test oracle for
ether-related transactions that examines the invariant between balances from a
bookkeeping variable and the ether values among accounts\cite{vultron}.


\noindent\textbf{Symbolic Execution:} Symbolic execution techniques~\cite{klee,
    exe, dart, diode} have been used to improve software security for many
    years.  KLEE~\cite{klee, exe} is a popular symbolic execution engine on
    LLVM framework for traditional computer programs. DIODE~\cite{diode}
    proposes goal-directed conditional branch enforcement technique to quickly
    explore execution paths to detect integer overflow errors. 

\section{Conclusion}\label{sec:conclusion}

We presented {\systemname}, a novel analysis tool for detecting standard
violation errors for Ethereum smart contracts. {\systemname} opeartes with an
optimized symbolic execution engine and utilizes the standardization effort of
the community. As a result, {\systemname} is significantly more effective than
previous tools in our experiments, finding more errors with much less false positives.
Furthermore, our results show that standard violation errors are prevalent even
in those critical smart contracts that manipulate digital assets.  This calls
for more community efforts to enforce the consistency between smart contract
implementations and contract standards.

{\normalsize \bibliographystyle{acm}
\bibliography{references}}

\clearpage
\appendix
\section{Error Triggering Input}\label{appendix:error-triggering}

\begin{figure}
\scriptsize
\begin{center}
\begin{tabular}{ |c|c| } 
\hline
Variable & Value \\ 
\hline
\sourcecode{ADDRS.length} & 3 \\ 
\sourcecode{ADDRS[0]} & 0x67518339e369ab3d591d3569ab0a0d83b2ff5198 \\ 
\sourcecode{ADDRS[1]} & 0xd30a286ec6737b8b2a6a7b5fbb5d75b895f62956 \\ 
\sourcecode{ADDRS[2]} & 0x1bfa530d5d685155e98cd7d9dd23f7b6a801cfef \\ 
\sourcecode{sender} & \sourcecode{ADDRS[0]} \\ 
\scriptsize{\sourcecode{balanceOf(sender)}} & \tiny{0x2cda3fa1fffffffe000000020008007e87fe00001c180f800046f383eb594355}\\ 
\sourcecode{\_receivers.length} & 3 \\ 
\sourcecode{\_receivers[0]} & \sourcecode{ADDRS[2]} \\ 
\sourcecode{\_receivers[1]} & \sourcecode{ADDRS[2]} \\ 
\sourcecode{\_receivers[2]} & \sourcecode{ADDRS[0]} \\ 
\sourcecode{\_value} & \tiny{0xb380b590a3ff87ff3ffff9560000c3baaaad1d4a74e77a85be86c5459cf3a343}\\
\hline
\end{tabular}
\end{center}
\caption{Concrete example exploits the \sourcecode{batchTransfer()} function.}
\label{fig:concrete}
\end{figure}

Figure~\ref{fig:concrete} shows a concrete error triggering transaction
{\systemname} generates for exploiting the error we described in
Section~\ref{sec:example}.

\section{Memory Operations}\label{appendix:memory}
Figure~\ref{code:opt} shows an example of storing and loading memory data without
any optimization. To store a 256-bit integer \sourcecode{value}, a naive approach would 
first generate 32 \sourcecode{extract} instructions and 32 \sourcecode{store} instructions.
Each pair of \sourcecode{extract} and \sourcecode{store} instructions 
extract a 8-bit integer from \sourcecode{value} and 
store the integer to the memory.
Similarly, to load a 256-bit integer from memory, 
the system would generate 32 \sourcecode{select}
instructions to load a 256-bit integer from the memory. 
Each instruction loads a 8-bit integer from memory. 
It then would concatenate 32 8-bit integers into a 256-bit integer.

\begin{figure}[t]
\begin{center}
\begin{minted}{Scheme}
(declare-fun MEMORY () (Array (_ BitVec 256) (_ BitVec 8)))

; MSTORE(offset,value) 
(declare-fun v_1 () (_ BitVec 8))(assert (= v_1 ((_ extract 7 0) value)))
(declare-fun s_1 () (Array (_ BitVec 256) (_ BitVec 8)))(assert (= s_1 (store Memory offset v_1)))
(declare-fun v_2 () (_ BitVec 8))(assert (= v_2 ((_ extract 15 8) value)))
(declare-fun s_2 () (Array (_ BitVec 256) (_ BitVec 8)))(assert (= s_2 (store s_1 offset+8 v_2)))
  ...
(declare-fun v_32 () (_ BitVec 8))(assert (= v_32 ((_ extract 248 255) value)))
(declare-fun s_32 () (Array (_ BitVec 256) (_ BitVec 8)))(assert (= s_32 (store s_31 offset+248 v_32)))

...

; MLOAD(offset)
(declare-fun a_1 () (_ BitVec 8))(assert (= a_1 (select s_32 offset)))
(declare-fun a_2 () (_ BitVec 8))(assert (= a_2 (select s_32 offset+8)))
  ...
(declare-fun a_32 () (_ BitVec 8))(assert (= a_5 (select s_32 offset+248)))
(declare-fun return () (_ BitVec 256))(assert (= return (concat a_1 a_2 ... a_32)))
\end{minted}
\end{center}
\caption{Constraints from MSTORE and MLOAD instructions without simplification.}
\label{code:opt}
\end{figure}

\section{Invariants and Constraints}\label{sec:constraints}

In this section, we describe how a user checks the 
invariants and the constraints using {\systemname}.

\begin{figure}[t]
\begin{minted}{Python}
# Get balances before transfer is called.
acc = [SymAddr(), SymAddr()]
assume(acc[0] != acc[1])
value = SymInt()
pre_bal = [c.balanceOf(account) for account in acc]
assume(pre_bal[0] + pre_bal[1] >= pre_bal[0])
# account1 transfers value tokens to account2.
result = c.transfer(acc[1], value, sender=acc[0])
# Check if security policies is violated.
post_bal = [c.balanceOf(account) for account in acc]
check(result == 0 and
      pre_bal[0] == post_bal[0] and
      pre_bal[1] == post_bal[1] or
      result != 0 and
      pre_bal[0] - value == post_bal[0] and
      pre_bal[1] + value == post_bal[1] and
      post_bal[0] >= pre_bal[0] and
      pre_bal[1] >= post_bal[1])
\end{minted}
\caption{Policies for transfer function.}
\label{code:transfer}
\end{figure}

\begin{figure}[t]
\begin{minted}{Python}
acc = [SymAddr(), SymAddr(), SymAddr()]
values = [SymInt(), SymInt()]
assume(acc[0] != acc[1] and acc[1] != acc[2])
# Get account balances before tokens are transferred.
pre_bal = [c.balanceOf(account) for account in acc[:2]]
assume(pre_bal[0] + pre_bal[1] >= pre_bal[0])
# acc[0] authorize acc[2] to spend values[0] tokens.
r1 = c.approve(acc[2], values[0], sender=acc[0])
assume(r1 != 0)
# acc[2] transfer values[1] tokens from acc[0]'s account to acc[1]'s account.
r2 = c.transferFrom(acc[0], acc[1], values[1], sender=acc[2])
r3 = c.allowance(acc[0], acc[2])
post_bal = [c.balanceOf(account) for account in acc[:2]]
# Check if the security policies are violated.
check(r2 != 0 and
      pre_bal[0] - values[1] == post_bal[0] and
      pre_bal[1] + values[1] == post_bal[1] and
      r3 + values[1] == values[0] and
      post_bal[0] <= pre_bal[0] and
      post_bal[1] >= pre_bal[1] and
      values[0] >= values[1] and
      values[0] <= r3 or
      r2 == 0 and
      pre_bal[0] == post_bal[0] and
      pre_bal[1] == post_bal[1] and
      r3 == values[0])
\end{minted}
\caption{Policies for approve and transferFrom functions (ERC-20).}
\label{code:approve}
\end{figure}

\begin{figure}[t]
\begin{minted}{Python}
acc = [SymAddr(), SymAddr()]
tid = SymInt()  # token id
assume(acc[0] != acc[1])
# get the owner address of the token.
owner = c.ownerOf(tid)
assume(owner != acc[1])
# assume the operator(acc[0]) is not authorized to
# manage the token.
app = c.getApproved(tid)
is_approved = c.isApprovedForAll(owner, acc[0])
assume(a[0] != app)
assume(is_approved == 0)
c.transferFrom(owner, a[1], tid, sender=acc[0])

# check if transfer is successful.
pos_owner = c.ownerOf(tid)
check(pos_owner == acc[1])
\end{minted}
\caption{Policies for approve and transferFrom functions (ERC-721).}
\label{code:approve-nft}
\end{figure}

\noindent{\textbf{Total Supply:}} As described in
Section~\ref{sec:example} and Figure~\ref{code:total},  ERC-20 tokens are
designed as assets that can be sent and received and in practice many 
ERC-20 tokens are traded publicly and have market value.  If the total supply changes
unexpectedly due to software errors, it may have significant impact on the
market value of each token.  We applied {\systemname} to check if any public
function in our benchmark contracts will break the total supply invariant.

\noindent\textbf{Transfer:} The \sourcecode{transfer()} function defines
the most fundamental functionality of an ERC-20 token, that allows a user to transfer his
balances to other users.  To secure a token transfer, the contract must
guarantee that the sender has enough token to transfer. Otherwise, the token
transfer should fail.

Figure \ref{code:transfer} shows the function to build transfer policy, 
where \sourcecode{acc[0]} and \sourcecode{acc[1]} are two symbolic accounts 
and \sourcecode{value} is a symbolic value. Before calling
the function \sourcecode{transfer()}, account \sourcecode{acc[0]} owns \sourcecode{pre\_bal[0]} tokens and \sourcecode{acc[1]} 
owns \sourcecode{pre\_bal[1]} tokens. \sourcecode{acc[0]} creates a transaction that transfers \sourcecode{value} 
tokens to \sourcecode{acc[1]}. \sourcecode{post\_bal[0]} and \sourcecode{post\_bal[1]} are tokens owned by \sourcecode{acc[0]}
and \sourcecode{acc[1]} after calling the \sourcecode{transfer()} function. At line 11-18, the generated
constraints check if the transfer function breaks the security rule defined above.

\noindent\textbf{Token Approval (ERC-20):} The \sourcecode{approve()} and
the \sourcecode{transferFrom()} functions allow a user to authorize a third party to
spend his token. A user calls the \sourcecode{approve()} function to allow a third
party to spend tokens in his account and the third party uses the
\sourcecode{transferFrom()} function to transfer the token.

Figure~\ref{code:approve} shows the security policies built for
\sourcecode{approve()} and \sourcecode{transferFrom()} scheme. At line 8 account
\sourcecode{acc[0]} calls the function \sourcecode{approve()} to authorize
\sourcecode{acc[2]} to spend up to \sourcecode{values[0]} tokens.
\sourcecode{acc[2]} then calls the \sourcecode{transferFrom} function to transfer
\sourcecode{acc[0]}'s token to \sourcecode{acc[1]}'s account.

A secure \sourcecode{transferFrom()} transaction should check the following
conditions. 1) \sourcecode{acc[0]} has enough tokens to perform the
transaction; 2) the transferred token is smaller than \sourcecode{acc[0]}'s
allowance; 3) the allowance and the account balances should be updated
correspondingly if the transfer succeeds (line 15-26).

\noindent\textbf{Token Approval (ERC-721):}
Similar to ERC-20, ERC-721 allows a user to transfer his ownership of a token
to other users.  Different from ERC-20, ERC-721 uses the \sourcecode{approve()} function to
authorize a third party to transfer a single token and
\sourcecode{setApprovalForAll} to authorize a third party to manage all
assets.

In Figure~\ref{code:approve-nft}, {\systemname} assumes that \sourcecode{acc[0]} is not
authorized to transfer token \sourcecode{tid} and \sourcecode{acc[1]} is not the owner
of \sourcecode{tid} (line 5-12). \sourcecode{acc[0]} then creates a transaction that transfers the ownership 
of \sourcecode{tid} from \sourcecode{owner} to \sourcecode{acc[1]} (line 13). A vulnerable contract would
allow the transaction and modify the ownership correspondingly.

\end{document}